\documentclass[sigconf]{acmart}

\AtBeginDocument{%
  \providecommand\BibTeX{{%
    \normalfont B\kern-0.5em{\scshape i\kern-0.25em b}\kern-0.8em\TeX}}}

\copyrightyear{2024}
\acmYear{2024}
\setcopyright{acmlicensed}
\acmConference[SIGIR '24]{Proceedings of the 47th
International ACM SIGIR Conference on Research and Development in
Information Retrieval}{July 14--18, 2024}{Washington, DC, USA}
\acmBooktitle{Proceedings of the 47th International ACM SIGIR Conference on
Research and Development in Information Retrieval (SIGIR '24), July 14--18,
2024, Washington, DC, USA}
\acmDOI{10.1145/3626772.3657717}
\acmISBN{979-8-4007-0431-4/24/07}
\usepackage{hyperref}
\usepackage{algorithm}
\usepackage{algorithmic}
\usepackage{subfigure}
\usepackage{amsthm}
\usepackage{color}
\usepackage{makecell}
\usepackage{multirow}
\usepackage{bm}
\usepackage{amsfonts}
\usepackage{url}
\usepackage{subfigure}
\usepackage{balance}
\usepackage{siunitx}
\usepackage{balance} 
\begin{document}

% \title{Law-aware Generative legal case retrieval}

% 标题需要确定
% \title{Incorporating Judgment Prediction into Legal Case Retrieval via Law-aware Generative Retrieval}
\title{Explicitly Integrating Judgment Prediction with \\ Legal Document Retrieval: A Law-Guided Generative Approach}

\author{Weicong Qin}
% \authornote{Both authors contributed equally to this research. Work partially done at Engineering Research Center of Next Generation Intelligent Search and Recommendation, Ministry of Education.}
\affiliation{%
   \department{Gaoling School of Artificial Intelligence}
  \institution{Renmin University of China}
  % \city{Beijing}
  % \country{China}
  \country{}
}
\email{qwc@ruc.edu.cn}

% \author{A. U. Thor}
% \orcid{1234-4564-1234-4565}
% \affiliation{%
% \institution{University of New South Wales}
% \department{School of Biomedical Engineering}
% \streetaddress{Samuels Building (F25), Kensington Campus}
% \city{Sidney}
% \state{NSW}
% \postcode{2052}
% \country{Australia}}
% \email{author@nsw.au.edu}

\author{Zelin Cao}
\affiliation{%
    \department{Gaoling School of Artificial Intelligence}
  \institution{Renmin University of China}
  % \city{Beijing}
  % \country{China}
  \country{}
}
\email{zelin_cao@ruc.edu.cn}

\author{Weijie Yu}
\affiliation{%
    \department{School of Information Technology and Management}
  \institution{University of International Business and Economics
}
  % \city{Beijing}
  % \country{China}
  \country{}
}
\email{yu@uibe.edu.cn}

\author{Zihua Si}
\affiliation{%
  \department{Gaoling School of Artificial Intelligence}
  \institution{Renmin University of China}
  % \city{Beijing}
  % \country{China}
  \country{}
}
\email{zihua_si@ruc.edu.cn}

\author{Sirui Chen}
\affiliation{%
  \institution{University of Illinois at Urbana-Champaign}
  % \city{Urbana Champaign}
  % \country{USA}
  \country{}
}
\email{chensr16@gmail.com}

\author{Jun Xu}
\authornote{Jun Xu is the corresponding author. Work partially done at Engineering Research Center of Next-Generation Intelligent Search and Recommendation, Ministry of Education.}
\affiliation{%
  \department{Gaoling School of Artificial Intelligence}
  \institution{Renmin University of China}
  % \city{Beijing}
  % \country{China}
  \country{}
}
\email{junxu@ruc.edu.cn}

% \affiliation{%

% \affiliation{%

% \affiliation{%

% \affiliation{%

% \affiliation{%

% \affiliation{%

% \affiliation{%

\renewcommand{\shortauthors}{Weicong Qin et al.}

% \newcommand{\ywj}[1]{{\color{violet} ywj: ``#1''}}
% \newcommand{\szh}[1]{{\color{orange} szh: ``#1''}}
% \newcommand{\csr}[1]{{\color{magenta} csr: ``#1''}}
% \newcommand{\qwc}[1]{{\color{blue} qwc: ``#1''}}
% \newcommand{\qwcC}[1]{{\color{violet} [\textbf{QWC's comments}: #1]}}
% \newcommand{\xujun}[1]{{\color{red} xujun: ``#1''}}

\begin{abstract}
Legal document retrieval and judgment prediction are crucial tasks in intelligent legal systems. In practice, determining whether two documents share the same judgments is essential for establishing their relevance in legal retrieval. However, existing legal retrieval studies either ignore the vital role of judgment prediction or rely on implicit training objectives, expecting a proper alignment of legal documents in vector space based on their judgments. Neither approach provides explicit evidence of judgment consistency for relevance modeling, leading to inaccuracies and a lack of transparency in retrieval. To address this issue, we propose a law-guided method, namely GEAR, within the generative retrieval framework. GEAR explicitly integrates judgment prediction with legal document retrieval in a sequence-to-sequence manner. Specifically, given the intricate nature of legal documents, we first extract rationales from documents based on the definition of charges in law. We then employ these rationales as queries, ensuring efficiency and producing a shared, informative document representation for both tasks. Second, in accordance with the inherent hierarchy of law, we construct a law structure constraint tree and represent each candidate document as a hierarchical semantic ID based on this tree. This empowers GEAR to perform dual predictions for judgment and relevant documents in a single inference, i.e., traversing the tree from the root through intermediate judgment nodes, to document-specific leaf nodes. Third, we devise the revision loss that jointly minimizes the discrepancy between the IDs of predicted and labeled judgments, as well as retrieved documents, thus improving accuracy and consistency for both tasks.  Extensive experiments on two Chinese legal case retrieval datasets show the superiority of GEAR over state-of-the-art methods while maintaining competitive judgment prediction performance. Moreover, we validate the effectiveness of GEAR on a French statutory article retrieval dataset, reaffirming its robustness across languages and domains.
\end{abstract}






\maketitle

\section{Introduction}
\label{sec:intro}
Legal document retrieval and judgment prediction are fundamental components in intelligent legal systems. The former entails the retrieval of relevant legal documents (cases or statutory articles) give a query. 
% an indispensable process in legal research, case analysis, and numerous other applications within the legal domain. 
On the other hand, the latter seeks to predict the outcomes or judgments rendered in legal cases, such as applicable charges, term-of-penalties, etc. 

\begin{figure}
    \centering
\includegraphics[width=0.82\linewidth]{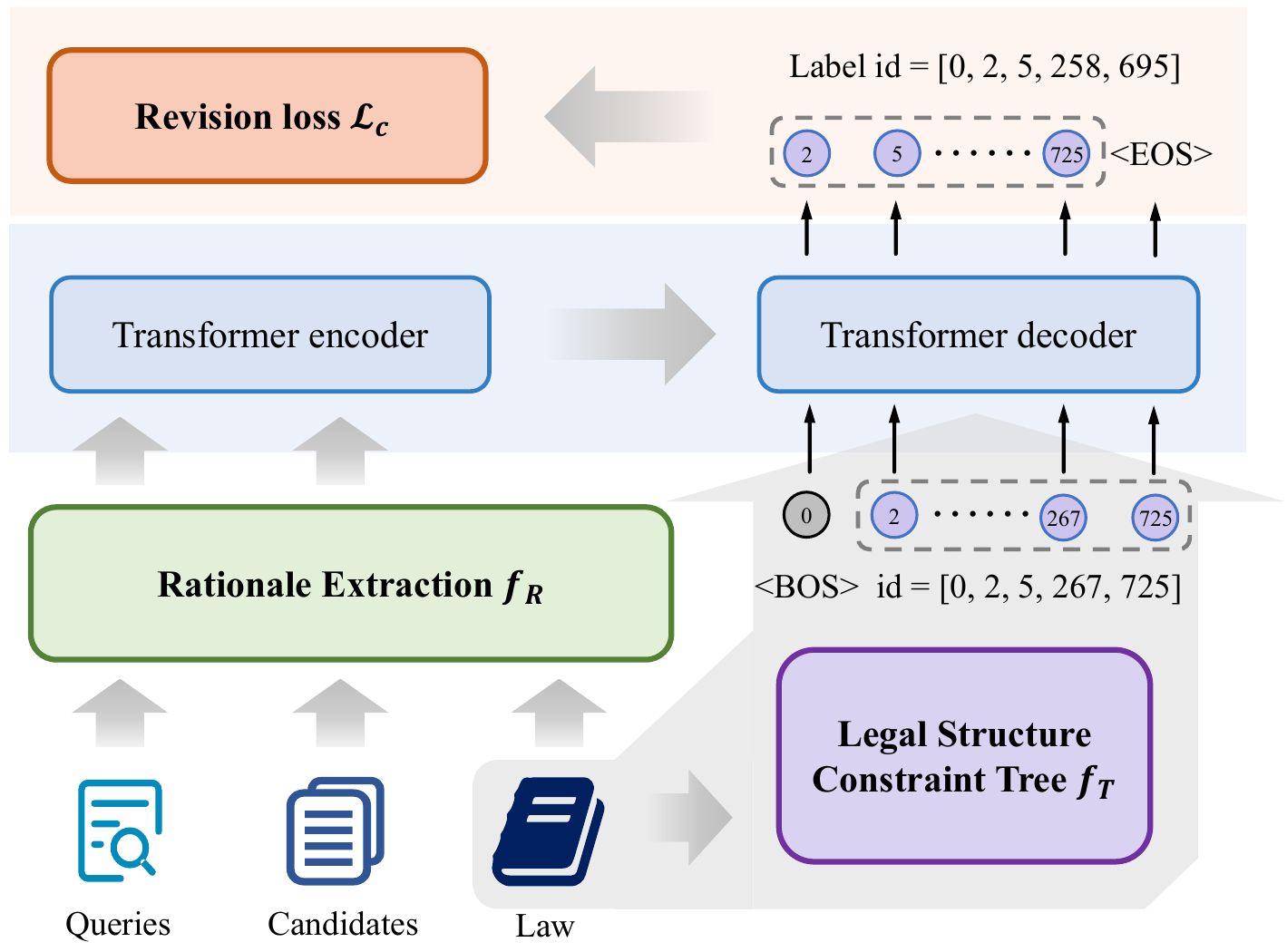}
    \caption{The overview of our proposed GEAR. It mainly consists of three modules, including rationale extraction, law structure constraint tree, and revision loss. The middle part in blue is the generative retrieval framework. }
    \label{fig:fig2}
% \vspace{-2.6ex}
\end{figure}

These two tasks are closely intertwined~\citep{shao2023intent,shao2023understanding,zhanghan} in practice. 
% Their interdependence is rooted in the intrinsic relationship between legal issues presented in cases and the judgments they ultimately yield.
From the retrieval side, determining whether two documents share the same judgments is essential for establishing their relevance. 
% In other words, without the evidence offered by judgment prediction, the retrieval process is often hampered, leading to the inclusion of irrelevant documents or the omission of vital ones. 
Regrettably, most of the existing studies~\citep{lecard,lawformer,chalkidis2020legal,sun2024logic,yu-etal-2022-optimal}  of legal document retrieval frequently overlook the significance of judgment prediction and merely focus on the text-level semantic similarity. 
Recently,~\citet{sailer} introduced an implicit training objective that uses the fact description of the legal document to predict its judgment, expecting a proper alignment of legal documents in vector space based on their judgments.  
While these studies show effectiveness in retrieval performance, they fail to provide explicit evidence of judgment consistency for relevance modeling. Consequently, this limitation leads to inaccuracies and a lack of transparency~\citep{olteanu2021facts,KIM2020113302,baan2019transformer}. It is because their legal relevance reasoning especially regarding judgment remains unclear, and we cannot trace back the decision-making process based on the retrieval results.
% the degree of relevance (e.g., how consistent the applicable judgments are) between the query and retrieved documents remains unclear, which in turn obscures the intricate legal reasoning process.
% \begin{figure}
%     \centering
% \includegraphics[width=0.9\linewidth]{figures/fig_1.pdf}
%     \caption{A pair of relevant cases from the LeCaRDv2 dataset. The applicable charges (underlined) and their corresponding rationales (red and blue) provide effective relevance signals. }
%     \label{fig:motivation}
% \end{figure}

Therefore, we aim to explicitly integrate judgment prediction with legal document retrieval. However, there remain the following challenges to achieve our goal.
% Despite the aforementioned benefits, 
% There remain the following challenges in incorporating judgment prediction into case retrieval.
% unifying both tasks. 
Firstly, legal document retrieval and judgment prediction are usually formulated as two distinct machine learning problems--retrieval and classification. It is difficult for one retrieval model to predict the applicable judgment for legal cases and in turn leverage the judgment prediction to enhance retrieval. 
Secondly, legal documents are inherently lengthy and complicated.
% As shown in~\autoref{fig:motivation}, only the rationales (key circumstances and key elements) ~\citep{lecard,yu2022explainable} of the case 
% related to law including key circumstances and key elements~\citep{lecard} 
% provide effective signals for relevance estimation. 
It results in the retrieval efficiency issue and hiders to represent each document as a shared and informative representation for both tasks.
% , but existing methods often encode legal cases in a coarse-grained granularity way~\citep{chalkidis2020legal,lawformer}.
Thirdly, both tasks rely on specialized law knowledge~\citep{sun2022law,10.1145/3539618.3591731},
an appropriate way that effectively injects law expertise to guide the prediction of both tasks remains a concern.
% given that current researches~\citep{chalkidis2020legal,lawformer,zhang2023contrastive,sailer} solely focus on text-level semantics for predictions.

% \csr{overall, this paragraph tends to treat two tasks equally. should the statement lean a bit to the main task, legal case retrieval?}

Facing the above challenges,
% and inspired by the success of generative retrieval~\cite{DSI, NCI}, 
we propose a novel law-guided generative legal document retrieval method, namely GEAR.
% , to incorporate judgment prediction into legal case retrieval.
% \csr{The connection between generative retrieval and the challenges is not clear. how about reorganize in this way: 1. generative retrieval shows potential to address this issue in a, b and c. 2. we combine this insight into the legal case retrieval.}
GEAR explicitly integrates judgment prediction with legal document retrieval in a sequence-to-sequence (Seq2Seq) manner as illustrated in~\autoref{fig:fig2}. 
% The insight of GEAR is to generate a set of law-aware semantic IDs where each ID represents a document relevant to the query and also reflects its applicable judgments.
The insight of GEAR lies in formulating the retrieval process into the generation of law-aware semantic IDs, where each ID not only represents a document relevant to the query but also reflects its applicable judgments.
% correspond to the relevant precedent cases of the query.
% This process mainly consists two steps: rationale extraction and decoding.
Specifically, we first construct a corpus based on the definition of charges in law\footnote{In this paper, we take laws in China and Belgium as examples. It is worth noting that GEAR can readily accommodate laws from various other countries.}
and subsequently extract rationales~\citep{yu2022explainable,sun2023explainable} representing the key elements/ circumstances~\citep{lecard,shao2023understanding} at the word- and sentence-level for each document according to this corpus. We employ these rationales instead of raw documents as queries for generation. 
% because it effectively filters out the noise to ensure efficiency and makes the rationales shared and informative to both tasks.
This strategy not only serves to effectively filter out the noise of legal documents, ensuring generation efficiency but also renders the rationales shared and informative for both tasks.
Then, we create the law structure constraint tree based on the inherent hierarchy of law (e.g. Chapter-Section-Article), considering that both tasks are learned with the guidance of law. Given this tree, we assign legal documents hierarchical semantic IDs, with the IDs reflecting their judgments, for example, the ID ``0-2-5-269-809'' indicates the document named 809 falls under Article 269 of Chapter 5 of Section 2.
% Thus, we can consider situations where each case may involve single or multiple charges and assign the law-aware semantic ID(s) to each case
% Besides, we consider single and multiple charges assign the law-aware semantic ID for each case 
% given this tree, making the ID reflect its charges.
In this way, the generation of these IDs is equivalent to traversing the tree from the root through intermediate judgment nodes, to document-specific leaf nodes. It makes GEAR capable of showing the legal reasoning process and performing dual predictions for judgment and relevant documents in a single inference.
% GEAR can show the legal reasoning process and perform dual predictions for judgment and relevant documents in a single inference, i.e., traversing the tree from the root through intermediate judgment nodes, to document-specific leaf nodes.
% To further improve the accuracy and consistency for both tasks, we devise the revision loss that jointly minimizes the discrepancy between predicted and labeled charges/ retrieved cases based on .
To further improve the accuracy and consistency of both tasks, we devise a novel training objective called the revision loss. This loss aligns with the hierarchy of the tree and jointly minimizes the discrepancy between predicted and labeled judgments/ retrieved cases.
% To further examine the performance on judgment prediction in case retrieval, we also propose a new metric called \textit{coverage} that indicates the charge-level consistency between the query and retrieved cases. 
% In the experiments, to assess the performance of judgment prediction in legal document retrieval, we introduce a novel metric called \textit{coverage}. This metric gauges the judgment-level consistency between the query and retrieved cases. 
Extensive experiments on two Chinese legal case retrieval datasets show the consistent superiority of GEAR over state-of-the-art methods while maintaining competitive judgment prediction performance. We also validate the effectiveness of GEAR on a French statutory article retrieval dataset, reaffirming its generalization ability.

% \begin{figure}
%     \centering
% \includegraphics[width=\linewidth]{figures/fig_1_v2.pdf}
%     \caption{
%   The architecture of our GEAR.
%     }
%     \label{fig:intro}
% \end{figure}

The major contributions of the paper are summarized as follows: 

(1) To the best of our knowledge, this is the first work that \textbf{explicitly} integrates judgment prediction with legal document retrieval. Our method is capable of showing the legal reasoning process and performing dual predictions for both tasks in a single inference. It improves the transparency of the legal decision-making.
% because our method can perform dual predictions for both tasks in a single inference. Our work improves the transparency of the legal decision-making process.

% (2) We propose a novel model, namely GEAR, to unify legal case retrieval and judgment prediction under the generative retrieval framework. We explicitly leverage the law knowledge to extract rationales from cases, assign cases the law-aware hierarchical identifies, and formulate the prediction as a traversal on the law structure constraint tree.
% We also propose a revision loss to jointly improve the accuracy and consistency of both tasks.

(2) We propose a novel law-guided generative model, namely GEAR. We explicitly leverage the law knowledge to extract rationales from legal documents, assign them the law-aware hierarchical IDs, and formulate the prediction as a traversal on the law structure constraint tree. We also propose the revision loss to jointly improve the accuracy and consistency of both tasks.

(3) We conduct extensive experiments on three public datasets of legal document retrieval in two languages. The results indicate GEAR not only achieves state-of-the-art performance in legal case and statutory article retrieval but also maintains competitive judgment prediction performance.

\section{Related Work}
\subsection{Legal Document Retrieval}
% Legal case retrieval is a special Information Retrieval (IR) task focusing on legal case documents. 
% 法律案例检索是一项专门的信息检索任务，目标是根据给定的查询的问题检索索相关的法律案例。与普通的检索任务相比，法律案例检索具有独特的特点。首先，由于司法调查的严谨性和全面性，法律案件文件通常比普通的文件长得多。同时，在实际司法程序中，查询往往是真实的待判决案例，也有很长的文本。不仅如此，法律案例检索的相关性定义超越了文本之间简单的语义相似性。这些特点带来了挑战从而使得普通的检索方法难以应用法律案例检索。在早期的几十年里，专家们投入了大量的努力来开发适合法律案例检索的方法，其中大致可以分为三类，早期的传统方法、中期的专业知识方法和近期的自然语言处理方法。
Legal document retrieval is a long-standing research topic in the field of information retrieval. 
% In contrast to general document retrieval, legal case retrieval involves a significant level of specialized legal knowledge and deals with lengthy and complicated legal documents.
In the early exploration, researchers~\citep{moens2001innovative,zeng2005knowledge,klein2006thesaurus,saravanan2009improving,bench2012history} made efforts to inject the legal knowledge to the retrieval through the decomposition of legal issues and involving the ontology.
In recent years, deep learning demonstrated its effectiveness in exploring document semantics. One representative line of works focused on the network-based precedents
methods tailored for the common law systems. For example,~\citet{minocha2015finding} leveraged the Precedent Citation Network (PCNet) to predict the relevance based on whether the sets of precedent citations occur in the same cluster.~\citet{bhattacharya2020hier} proposed Hier-SPCNet to capture all domain information inherent in laws and precedents.
The other line of works judged the relevance between the query and candidate document according to their text-level similarity. 
One such method was BERT-PLI~\cite{shao2020bert}. 
It divided the legal document into several paragraphs and used BERT~\citep{devlin2018bert} to obtain the similarity between the paragraphs.
% , achieving promising performance.
% It first split each case into paragraphs and models the interactions between query and candidate at the paragraph-level. Then a Pre-trained language models~\cite{devlin2018bert} was applied to encode every paragraph in the paired  cases, followed by a max-pooling operation for capturing the local relevance signals. Finally, a recurrent neural network (RNN) with an attention mechanism was used to predict the final relevant score.
Lawformer~\citep{xiao2021lawformer} was another text-based method. It used millions of Chinese criminal and civil case documents to pre-train a  Longformer~\citep{beltagy2020longformer} model.
% and achieved impressive performance across legal tasks.
Despite impressive performance, these works overlook the significance of judgment prediction and merely focus on the text-level similarity, resulting in sub-optimal and unreliable results.

Recently,~\citet{sailer} introduced an implicit training objective that uses the fact description of the legal document to predict its judgment, expecting a proper alignment of legal documents in
vector space based on their judgments. It fails to provide explicit evidence of judgment consistency for relevance modeling, leading to inaccuracies and a lack of transparency.

\subsection{Generative Retrieval}

% \szh{Give a simple example. Need to be extended.}
Generative retrieval has recently emerged as a promising direction for document retrieval. These methods assign semantic IDs to documents and utilize language models for ID generation. It enables an end-to-end retrieval in contrast to the traditional index-then-retrieve paradigm.
Various methods have been introduced to generate semantic document IDs. For example, ~\citet{cao2021autoregressive} introduced a  Seq2Seq system to conduct entity retrieval. They first represented documents as unique names that are composed of entity names. Then they used a auto-regressive generation model to generate the unique names of these entities based on contextual information. 
~\citet{DSI} proposed the differentiable search index (DSI) paradigm, which is an auto-regressive generation model to perform ad-hoc retrieval tasks. The input of the model was a natural language query and the model regressively generated documents’ ID strings that are relevant to the given query. ~\citet{NCI} proposed a novel method NCI, which used a tailored prefix-aware weight-adaptive decoder to optimize the retrieval  performance. 
% NCI also used constrained beam search to make sure all the generated IDs were valid.
Ultron~\cite{zhou2022ultron} leveraged document titles and substrings as IDs to enrich the semantic information of IDs.
To mitigate data distribution mismatch that occurs between the
indexing and the retrieval phases,
~\citet{DSI-QG} proposed DSI-QG, which adopted a query generation model with a cross-encoder to generate and select a set of relevant queries.

Existing studies focused on the general domain, lacking specific designs for legal documents and the integration of law knowledge.

\section{Methodology}
\subsection{Task Formulation}
In this work, we target on legal document retrieval including legal case retrieval (LCR) and statutory article retrieval (SAR).
Suppose that we have a set of collected samples $\mathcal{D}=\{(q, \mathcal{C}, \mathcal{R})\}$. For each data instance, $q$ is the query representing an undecided legal case submitted by the legal practitioner in LCR, a legal question in SAR; $\mathcal{C}=\{c_1, c_2, \cdots, c_{N}\}$ with size $N\in\mathbb{N}^+$ is the candidate precedent case set in LCR, the statutory article pool in SAR;  $\mathcal{R}$ represents the labeled relevant case/ statutory article set from $\mathcal{C}$ given the query $q$. 
% Legal case retrieval task requires to retrieve $\mathcal{R}$ from $\mathcal{C}$ given $q$.
Unlike previous studies that only predict to retrieve $\mathcal{R}$ from $\mathcal{C}$ given $q$, in this work, we instead unify judgment prediction and document retrieval into a generative retrieval framework, and thus aim at learning a retrieval function $f: q\times \mathcal{C}\rightarrow \mathcal{R}\times \mathcal{E}$, where $\mathcal{E}$ denotes the set of applicable judgment corresponding to $q$. 

% The proposed GEAR aims at incorporate legal judgment prediction  into case retrieval through following three modules:
% (1);(2);(3).

\subsection{Overall Framework}
% \subsubsection{Generative Retrieval}
To learn $f$ and explicitly integrate judgment prediction with legal document retrieval, 
% To retrieve relevant legal documents and predict their applicable judgment, 
we develop GEAR, a novel law-guided generative approach from the viewpoint of generative retrieval.
Essentially, given a query document, GEAR adopts a language model to perform the Seq2Seq generation where the retrieved documents are represented as semantic IDs.
Following the practice of~\citep{DSI,NCI,DSI-QG,zhou2022ultron},
GEAR consists of two major steps to directly generate IDs of documents as the retrieval target. 
In the first indexing step that focuses on memorizing the information about each document, our
GEAR takes each document $c$ as input and generates its ID ${id}^c$ as output. 
% in a straightforward Seq2Seq fashion.
The model is trained with the standard language model objective with the teacher forcing:
\begin{equation}
% \vspace{-0.4ex}
\label{eq:index}
\mathcal{L}_{i}=\sum_{c\in\mathcal{C}}\log P({id}^c|c).
% \vspace{-0ex}
\end{equation}
% where ${id}^c$ is represented by the atomic ID or the semantic ID. 
% Atomic IDs are generated by assigning each document an arbitrary integer, while semantic IDs are generated according to the hierarchical clustering algorithm employed over all the documents.
In the second retrieval phase, GEAR models associate each $q$ to its relevant document $r$ through an auto-regressive generation:
\begin{equation}
\label{eq:retrieval}
\mathcal{L}_{r}=\sum_{(q,r)\in\mathcal{D}}\log P({id}^r|q),
\end{equation}
where ${id}^r$ denotes the ID of $r$. 
% ${id}^r$ is also represented by the atomic ID or the semantic ID. 
As such, once a GEAR model is trained, it can be used to retrieve candidate documents for a test query in an end-to-end manner using beam search.

% In the legal scenario, existing GR methods
% both $q$ and $c$ are extremely lengthy documents, directly feed the document  inevitably confuses the model and hinders the learning efficiency. Moreover, 
As aforementioned in Section~\ref{sec:intro}, 
% applying GR to our legal scenario faces several challenges 
there are several challenges in the legal domain.
For explicitly integrating judgment prediction with legal document retrieval within GEAR, the critical learning tasks become:
(1) extract rationales instead of using raw documents to form the input for generation (\textbf{Section~\ref{sec:extraction}}); 
(2) create informative law-aware IDs for each document based on the hierarchical structure of law (\textbf{Section~\ref{sec:id}});
(3) develop a training objective to explicitly ensure judgment-level and document-level consistency between predictions and labels (\textbf{Section~\ref{sec:training}}).
\begin{figure*}[t]
    \centering
\includegraphics[width=0.85\linewidth]{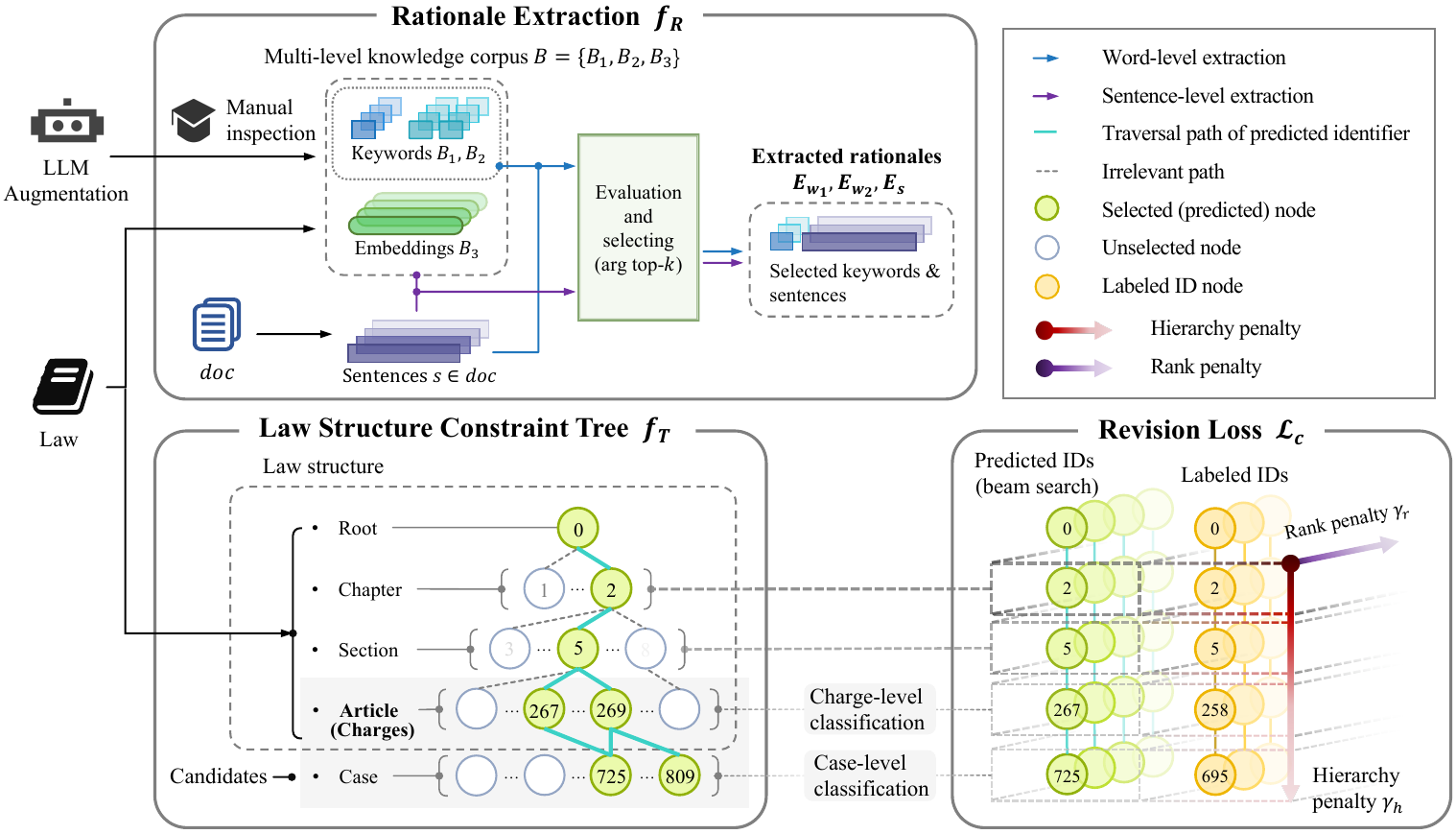}
    \caption{The proposed three modules of GEAR. $f_R$ extract rationales from legal documents; $f_T$ assigns hierarchical IDs to each document and constrain the decoding; $\mathcal{L}_c$ jointly optimizes document retrieval and judgement prediction.}
    \label{fig:fig_3}
\end{figure*}

\subsection{Rationale Extraction}
\label{sec:extraction}

% \qwc{(mainly show motivation in intro). According to previous research\citep{yu2022explainable,sun2023explainable}, we find that rationale plays a crucial role in the matching of legal documents. Simultaneously, considering the lengthy nature of legal documents (like cases) reduces efficiency, we decide to extract rationales\footnote{Following existing studies in the legal domain \citep{yu2022explainable,sun2023explainable} , "rationales" in our paper is defined as the subset of the document sentences that contributes to the final prediction. }. Additionally, we have identified a scarcity of large-scale sentence-level annotations in the legal retrieval scenario. Moreover, we have only found datasets with document-level charges annotations. Consequently, we use prototypes derived from legal charges to perform weakly supervised filtering and extraction of sentence-level retionales. }
% % Judicial case documents contain very long texts, even for queries, because the text that serves as a query in an actual judicial process is often the fact part of a case that has not yet been adjudicated and occupies the main part of the case text. Therefore, we believe that in order to bridge the gap between excessively long texts and model input constraints, it is necessary to extract the key elements for both historical and query case texts.

To ensure efficiency and provide shared and informative representations for legal documents, based on the set of laws $L$, we devise a module called $f_R$ to extract rationales $E$ instead of using the raw document $doc\in\{q,c\}$ as the input of GEAR:
\begin{equation}
    E = f_R(doc, L),
\end{equation}
where $E=\{E_w, E_s\}$ in which $E_w$ and $E_s$ respectively denotes the word-level and sentence-level rationales. 

As illustrated in~\autoref{fig:fig_3}, we first leverage the guidance of law to construct a corpus $B=\{B_1, B_2, B_3\}$, collecting law-based keywords $B_1, B_2$ and embeddings $B_3$. The keyword set
$B_1$ is constructed according to the lexical variants of all charge names in $L$. For example, in terms of the charge ``crime of forges the seals of a company, enterprise, institution or a people's organization''(translated from Chinese), we split the charge name and remove the stop words, then add the rest and their lexical variants to $B_1$.
% and put all keywords, \qwc{remaining words after removing stopwords}, and their lexical variants to the corpus. \qwc{We consider it as the golden keyword corpus.}
The keyword set $B_2$ is constructed similarly to the $B_1$.
We split the definitions of charges in law and remove the stop words then add the rest and their lexical variants to $B_2$. For the augmentation purpose, 
we employ the definitions of charges in law as prompts for the large language model\footnote{We use ChatLaw~\citep{cui2023chatlaw} for Chinese data. Since we have not found a suitable LLM for the Belgian legal domain, we omit this step for this data.}  (LLM) designed for the legal domain. After obtaining feedback from LLM, we remove stop words and incorporate the results into $B_2$.
To avoid the hallucination issue, we engage legal experts to manually assess the quality of augmentations to ensure the effectiveness of $B_2$.
% We put all keywords in the definition of laws and the feedback from prompting 
% \qwc{Large Language Models (LLM) in Legal Aspect}
% \footnote{\qwc{For Chinese legal domain, we adopt ChatLaw~\citep{cui2023chatlaw} (\url{https://github.com/PKU-YuanGroup/ChatLaw}), the current state-of-the-art LLM for this domain. For Belgian legal domain, we have not found a suitable LLM, so we omit this step.}} to the corpus after the stop words are removed. 
% Please note that while ChatLaw is a specialized large language model tailored for the legal domain, we have engaged legal experts to manually assess the quality of augmentations.
% \qwc{We have engaged legal experts to manually assess the quality of LLM augmentations to ensure the effectiveness of $B_2$, considering it as the universal keyword corpus. }
$B_3$ is constructed by collecting Legal-BERT~\citep{devlin2018bert,chalkidis2020legal} embeddings for the definition of each charge in law. 

Once the corpus is collected, we compute the multiple-level scores and extract $E$ for each document based on three corpora as follows. At the word level, we split the document and respectively select the top-$k_1$  (top-$k_2$) keywords $E_{w_1}$  ($E_{w_2}$) from $B_1$ ($B_2$) as follows:
\begin{equation}
\label{eq:word}
    % E_{w_i} = \arg\text{top-}k_i(\text{tf}( B_i,doc)),
    E_{w_i} = \mathop{\arg\text{top-}k_i}_{w\in B_i}\biggl(\text{tf}(w, doc)\biggr),
\end{equation}
where $i\in\{1,2\}$; $w$ is the word from $B_i$; $k_1$ and $k_2$ are hyperparameters to control the number of selected words; tf($\cdot$) denotes term frequency. At the sentence level, for each sentence $s$ in $doc$, we first select $E_s$ from as follows: 
\begin{equation}
\begin{aligned}
% E_s=\arg\text{top-}k_3\biggl(&\lambda_1 \frac{\text{sim}(s,B_1)}{\text{len}(doc)}+\lambda_2\frac{ \text{sim}(s,B_2)}{\text{len}(doc)}+\\
E_s=\mathop{\arg\text{top-}k_3}_{s\in doc, l\in L}\biggl(&\lambda_1 \frac{\text{sim}(s,B_1)}{\text{len}(s)}+\lambda_2\frac{ \text{sim}(s,B_2)}{\text{len}(s)}+\\
&\lambda_3\cos(\text{emb}(s),\text{emb}(l))\biggr),
\end{aligned}
\end{equation}
where $k_3$ is another hyperparameter to control the number of selected sentences; $\lambda_1, \lambda_2, \lambda_3$ are balance coefficients; len() denotes the sentence length; $\cos(\cdot)$ denotes the cosine similarity; emb() denotes the embedding function; sim($\cdot$) is defined as:
\begin{equation}
    \text{sim}(s, B_i)=\sum_{w\in s}\text{tf}(w,B_i),
\end{equation}
where $w$ is the word from $s$.
Please note that since query documents are typically undecided i.e., without labeled applicable charges, we extract rationales using the method described above. 
For the candidate precedent documents, whose applicable charges are given, we shrunk $B$ into a corpus constructed based on their corresponding labeled charges.

\subsection{Law Structure Constraint Tree}
\label{sec:id}
Given that both legal document retrieval and judgment prediction require guidance by law, typically organized in a ``Chapter-Section-Article'' hierarchy,
%  Considering that both legal document retrieval and judgment prediction require guidance by law where there usually exists a ``Chapter-Section-Article'' hierarchy, 
we argue that the decision process to judge whether two documents are relevant in document retrieval is analogous to the search in such a tree-like hierarchy. In other words, when legal practitioners search for the relevant documents given a query, they always expect the charge of relevant documents to be located at the same position in the law hierarchy as the charges applicable in the query.
Therefore, we devise a module $f_T$ that leverages the inherent hierarchy of the law to construct a law structure constraint tree $T$ as illustrated in~\autoref{fig:fig_3} and assigns the law-aware semantic ID $id$ for each $c\in\mathcal{C}$:
\begin{equation}
    id = f_T(c, T),
\end{equation}
where 
$id$ is in the prefix-suffix style.
The prefix depends on the position of the applicable charges within the tree.
As shown in~\autoref{fig:fig_3}, 
document 809 involves the charge of ``crime of robbery'' which falls under Article 269 of Chapter 5 of Section 2 of the Criminal Law of the People's Republic of China. Hence, the assigned prefix for this document is 0-2-5-269, where 0 represents the root node of the tree. As for the suffix, we regard documents are the children of their corresponding charges on the tree, and assign a unique ID to each of them under a crime node. Compared to current works that employ hierarchical $k$-means to create IDs for each document, ours avoid same integers may have different meanings at different levels, and thus ensure the effectiveness of model training.

% \qwc{During retrieval (inference)}, the IDs of top relevant \qwc{documents} can be easily generated via beam search. Due to the hierarchical nature of laws as analyzed above, it is convincing to use constrained beam search~\citep{anderson2017guided,hu-etal-2019-improved,post2018fast} decoding process with our law structure constraint tree, which \qwc{constrains the search space and} in turn only generates valid IDs.

% \begin{algorithm}
% \caption{Beam Search Generation Model}
% \begin{algorithmic}[1]
% % \State Initialize beam search with $k$ beams:
% %   \State \quad Set $B_0 = \{ \text{<start>} \}$, where $|B_0| = 1$
% % \For {$t = 1$ to $T$}
% %   \State Score each partial hypothesis $h$ in $B_{t-1}$ using the model:
% %     \State \quad Compute $\text{score}(h)$ for each $h \in B_{t-1}$
% %   \State Select the top $k$ hypotheses to form $B_t$:
% %     \State \quad $B_t = \text{Top-}k\left(\bigcup_{h \in B_{t-1}} \text{Expand}(h)\right)$
% % \EndFor
% % \State Select the final hypothesis $\hat{h}$ from $B_T$ with the highest score:
% %   \State \quad $\hat{h} = \text{argmax}_{h \in B_T} \text{score}(h)$
% % \State Output $\hat{h}$ as the generated sequence.
% \end{algorithmic}
% \end{algorithm}

On the other hand, one unique feature of the legal domain is that a single document can involve multiple charges. When the query is this kind of document, the ideal retrieval results should encompass these charges. Therefore, we assign $k$ IDs to documents involving $k$ charges, with each ID corresponding to a specific charge. For example, as shown in~\autoref{fig:fig_3}, the 725 involves charge 267 and charge 269, the valid IDs for this document include 0-2-5-267-725 and 0-2-5-269-725. In doing so, during retrieval, we expect the model to retrieve all IDs of the target documents, thereby increasing the probability of the target being retrieved. 

% \begin{align}
% & \text{Initialize beam search with } k \text{ beams:} \nonumber \\
% & \quad \text{Set } B_0 = \{ \text{<start>} \}, \text{ where } |B_0| = 1 \nonumber \\
% & \text{for } t = 1 \text{ to } T \text{ do:} \nonumber \\
% & \quad \text{Score each partial hypothesis } h \text{ in } B_{t-1} \text{ using the model:} \nonumber \\
% & \quad \quad \text{Compute } \text{score}(h) \text{ for each } h \in B_{t-1} \nonumber \\
% & \quad \text{Select the top } k \text{ hypotheses to form } B_t: \nonumber \\
% & \quad \quad B_t = \text{Top-}k\left(\bigcup_{h \in B_{t-1}} \text{Expand}(h)\right) \nonumber \\
% & \text{end for} \nonumber \\
% & \text{Select the final hypothesis } \hat{h} \text{ from } B_T \text{ with the highest score:} \nonumber \\
% & \quad \hat{h} = \text{argmax}_{h \in B_T} \text{score}(h) \nonumber \\
% & \text{Output } \hat{h} \text{ as the generated sequence.} \nonumber
% \end{align}

% \vspace{-1ex}
\subsection{Revision Loss}
\label{sec:training}
Besides employing the typical language model training objectives (\autoref{eq:index} and~\autoref{eq:retrieval}), we also develop a novel training objective called the revision loss for the consistency between the query and retrieved documents, i.e., we aim to directly  minimize their judgment-level and document-level discrepancy.

% We view the problem of case retrieval from the perspective of reinforcement learning. Specifically, We formulate the traversal on the law-aware decoding tree, i.e., retrieval, as a Markov Decision Process (MDP), which consists of $5$-tuple
% $\langle\mathcal{S}, \mathcal{A}, \mathcal{P}, \mathcal{R}, \gamma\rangle$: 
% $\mathcal{S}$ is the discrete state space, where $s\in \mathcal{S}$ indicates the state of a user including static features such as gender and dynamic features such as historical interactions.

Formally, as illustrated in~\autoref{fig:fig_3}, given the predicted ID (list of integers) $[\hat{id}_1, \hat{id}_2,\cdots, \hat{id}_L]$ and the corresponding ground-truth ID $[{id}_1, {id}_2,\cdots, {id}_L]$ for a query $q$, both having a length of $L$, we compare the difference between them and calculate the reward $R_t$ at each step $t\in[1,L]$ as follows:
% \begin{equation}
% \label{eq:reward}
% R_t = \gamma^{t}|i^p_t-i^g_t|,
% \end{equation}
\begin{eqnarray}
R_t =
\begin{cases}
\mu,  & \text{if}~\hat{id}t={id}_t,\\
-\gamma_h^{L-t}\mu|\hat{id}_t-{id}_t|, & \text{if}~\hat{id}_t\neq {id}_t,\\
\end{cases}
\end{eqnarray}
where $\mu$ is the constant reward unit,
$\gamma_h\in(0,1]$ is the hierarchy penalty factor used to penalize the differences between predictions and labels layer by layer along the law structure constraint tree, with larger penalties as it gets closer to the tree root and smaller penalties as it gets closer to the tree leaves.
Intuitively, if two documents share the same prefix, they are likely relevant to each other because of the same applicable charge, receiving higher rewards. 
% Conversely, if two documents belong to different chapters in terms of applicable charges, they are likely dissimilar resulting in lower rewards.

Then we apply the REINFORCE algorithm~\citep{williams1992simple} to optimize the model parameters, the revision loss is defined to minimize the policy gradient objective:
\begin{equation}
\label{eq:rl}
\mathcal{L}_c=-\sum_{q\in\mathcal{D}}\sum^{bz}_b\gamma_r^{b}\sum^{L}_t\log p(\hat{id}_t|q)\cdot R_t - \lambda_w\log p({id}_t|q),
\end{equation}
where $\gamma_r\in(0,1]$ the optional penalty factor used to focus on the top of retrieved document list; $bz$ denotes the beam size, i.e., the number of documents to retrieve for each query; $p(\hat{id}_t|q)$ is the probability that predicting to generate $\hat{id}_t$ given $q$ at layer $t$; $\lambda_w$ is the balance coefficient.
To further handle the sparse reward issue and improve the training efficiency, we follow~\citep{xin2020self,chen2021user,liu2023exploration} and add the second term   to~\autoref{eq:rl} that directly increases the probability of generating ${id}_t$.
Thus, the overall training objective is:
\begin{equation}
    \mathcal{L} = \mathcal{L}_i + \mathcal{L}_r + \lambda_l\mathcal{L}_c,
\end{equation}
where $\lambda_l$ is the coefficient to balance the indexing loss~(\autoref{eq:index}), retrieval loss~(\autoref{eq:retrieval}), and the revision loss. 

\subsection{Inference}
In the inference, we aim to retrieve the top-$k$ documents from the candidate pool. Since we have assigned hierarchical semantic ID to each document based on the law-aware constraint tree where each leaf node corresponds to a candidate document in the pool, we utilize the constrained beam search~\citep{anderson2017guided,hu-etal-2019-improved,post2018fast} to ensure all the generated document IDs are valid within the tree. 

\section{Experiments}
In this section, we conduct experiments to answer the following research questions: 
% In this section, we conduct experiments\footnote{The source code, datasets, and all experiments have been shared at:~\url{https://anonymous.4open.science/r/GEAR-www}.} to answer the following research questions: 
% \\
% (1) \textbf{RQ1:} How does GEAR perform on legal case retrieval compared to existing methods?\\
% (2) \textbf{RQ2:} How effective are the three modules in GEAR?\\
% (3) \textbf{RQ3:} Can GEAR show considerable performance on judgment prediction? \\
% (4) \textbf{RQ4:} What is the quality of the rationales extracted by GEAR including effectiveness and efficiency? \\
% (5)\textbf{RQ5:} Can GEAR incur less time overhead in legal case retrieval compared to popular generative methods?
% \begin{itemize}
% \item \textbf{RQ1:} How does GEAR perform on legal document retrieval compared to state-of-the-art methods?
% \item \textbf{RQ2:} How effective are the three modules in GEAR?
% \item \textbf{RQ3:} Can GEAR show competitive performance on judgment (applicable charges) prediction?
% \item \textbf{RQ4:} What is the quality of the rationales extracted by GEAR including effectiveness and efficiency?
% \item \textbf{RQ5:} Can GEAR incur less time overhead in legal document retrieval compared to popular generative methods?
% \item \textbf{RQ6:} How robust is GEAR across languages and domains (e.g. in statutory article retrieval)?
% \end{itemize}
\textbf{RQ1:} How does GEAR perform on legal document retrieval compared to state-of-the-art methods? \textbf{RQ2:} How effective are the three modules in GEAR? \textbf{RQ3:} Can GEAR show competitive performance on judgment (applicable charges) prediction? \textbf{RQ4:} What is the quality of the rationales extracted by GEAR including effectiveness and efficiency? \textbf{RQ5:} Can GEAR incur less time overhead in legal document retrieval compared to popular generative methods? \textbf{RQ6:} How robust is GEAR across languages and domains (e.g. in statutory article retrieval)?

% \textbf{RQ1:} How does GEAR perform on legal case retrieval compared to state-of-the-art methods? 
% \textbf{RQ2:} How effective are the three modules in GEAR? \textbf{RQ3:} Can GEAR show competitive performance on judgment (applicable charges) prediction? 
% \textbf{RQ4:} What is the quality of the rationales extracted by GEAR including effectiveness and efficiency? 
% \textbf{RQ5:} Can GEAR incur less time overhead in legal case retrieval compared to popular generative methods? 
% \textbf{RQ6:} How robust is GEAR across languages and domains(e.g. in statutory article retrieval)?

\subsection{Experimental Settings}
\label{sec:setting}
% \subsubsection{Datasets}
% We conduct experiments on  LeCaRDv2\citep{lecard} and ELAM\citep{yu2022explainable} for legal case retrieval.~\autoref{tab:dataset statistics} reports basic statistics of both datasets. We leave the detailed  descriptions in Appendix~\ref{sec:app:data}.
\subsubsection{Datasets}
\label{sec:app:data}
% We conduct experiments on ELAM~\citep{yu2022explainable} and LeCaRDv2~\citep{lecard} for legal case retrieval, on BSARD~\citep{BSARD} for statutory article retrieval.
% % ~\citep{cha}

% To answer \textbf{RQ1-5},
% We first assess GEAR's capability on two Chinese LCR datasets.

\textbf{ELAM}\footnote{\url{https://github.com/ruc-wjyu/OPT-Match}.}~\cite{yu2022explainable} is a Chinese LCR dataset, focusing on criminal cases. 
ELAM has corresponding labels for both case retrieval and judgment prediction, which is suitable for our goal.
We exclude those cases with multiple applicable charges to consider the retrieval and judgment prediction performance in 
a single charge scenario.
The resulting candidate pool size of ELAM is 1332. Other data preprocessing is aligned to~\cite{sailer}.

\textbf{LeCaRDv2}\footnote{\url{https://github.com/THUIR/LeCaRDv2}.}~\citep{li2023lecardv2} is the official updated version of LeCaRD ~\cite{lecard}. 
In this dataset, the relevance labels are divided into four levels, ranging from 3 to 0, indicating a gradual decrease in relevance. We follow the data preprocessing approach of~\cite{sailer}, with the exception of increasing the candidate pool size from 100 to 1390 to further validate the effectiveness of baselines and our model. In LeCaRDv2, cases encompass both single and multiple charges, averaging 1.5 charges per case.
% On LeCaRDv2, we consider both single and multiple charges, with an average of 1.5 charges per case.
Considering the ground-truth judgment labels of query cases have not been provided, we ask two legal experts (Ph.D. in Law) to annotate the charge label for the testing queries. The experts are proficient in Chinese criminal law with sufficient experience in handling cases similar to this dataset. They carefully align the LeCaRDv2 judgment criteria~\citep{li2023lecardv2} before annotation and discuss opinions to reach a consensus, ensuring accurate labeling. 
% Considering that the ground-truth judgment labels of query cases on LeCaRDv2 have not been provided so far, we ask two legal experts (Ph.D. in Law) to annotate the charge label for the testing queries. The experts is proficient in Chinese criminal law and has experience in handling cases similar to this dataset. Before annotation, they carefully read LeCaRDv2 judgment criteria to ensure accurate labeling. During the annotating, they discuss opinions and reach a consensus to provide consistent results.

% Although LeCaRDv2 does not provide the judgment labels, we instead create pseudo labels using the proposed metric $coverage$ in Section~\ref{sec:setting}.
% To answer \textbf{RQ6}, we conduct experiments on a Belgium SAR dataset in French to further evaluate the robustness of GEAR across languages and domains. 

\textbf{BSARD}\footnote{\url{https://huggingface.co/datasets/maastrichtlawtech/bsard}.}~\citep{BSARD} is a SAR dataset composed of more than 1.1K legal questions labeled by domain experts with relevant articles selected from the 22K law articles gathered from 32 publicly available Belgian codes. It is worth noting that BSARD contains structural annotations of corresponding laws, facilitating the utilization of law structural knowledge. 
% To validate the generalization capability of GEAR, we use BSARD dataset to evaluate GEAR's effectiveness in a distinct legal document retrieval task (statutory article retrieval) within the legal system of another country.  

\subsubsection{Baselines}
\label{sec:app:baseline}
We consider three types of baselines in this study.

\textbf{(1) Sparse retrieval methods:}\ \textbf{Query Likelihood (QL)}~\cite{zhai2008statistical} 
is a probabilistic language modeling approach employed to assess the relevance of documents to a provided query. 
% In QL, the likelihood of a document generating the words in a query is computed, treating both the query and the document as stochastic variables. 
% This methodology operates under the presumption that documents of relevance are inherently more inclined to generate the query terms.
% is a probabilistic language modeling approach used to estimate the relevance of documents to a given query. In QL, the likelihood of a document generating the words in a query is calculated, treating both the query and the document as random variables. This approach is based on the assumption that relevant documents are more likely to generate the query terms.
\textbf{BM25}~\cite{robertson2009probabilistic} is a probabilistic information retrieval model widely used in the field of text retrieval. BM25 takes into account both term frequency and document length normalization.
% , making it a robust and effective retrieval model.
% (1) \textbf{QL}~\cite{zhai2008statistical} is a representative traditional retrieval model based on Dirichlet smoothing;\\
% (2) \textbf{BM25}~\cite{robertson2009probabilistic} is a classic sparse retrieval model based on inverted-index.\\ 

\textbf{(2) Dense retrieval methods:} 
% \textbf{ANCE}~\cite{xiong2020approximate} is a contrastive learning based model for ad-hoc dense retrieval. It selects hard training negatives from the entire corpus to reduce the stochastic gradient
% variance, and improve retrieval performance.
\textbf{BERT}~\cite{devlin2018bert} is a strong baseline in ad-hoc retrieval tasks in the open domain. In this paper, we adopt the checkpoint that is pre-trained on a large Chinese corpus
%\footnote{\url{https://huggingface.co/bert-base-chinese}.}.
Following~\citep{DSI,NCI,DSI-QG}, after encoding legal documents using BERT, we then apply Approximate Nearest Neighbor (ANN) search algorithms to retrieve relevant documents. 
\textbf{Legal-BERT}\footnote{\url{https://github.com/thunlp/OpenCLaP}.}~\cite{chalkidis2020legal} is a variant of BERT that undergoes specific training in the legal domain to better understand and process text related to law. 
% In this work, we adopt the Legal-BERT pre-trained on a large Chinese legal corpus.
\textbf{Lawformer}
%\footnote{\url{https://huggingface.co/thunlp/Lawformer}.}
~\cite{lawformer} is a Longformer~\citep{beltagy2020longformer} backbone pre-trained on large legal case corpus, to encode legal texts.
\textbf{ChatLaw-Text2Vec}\footnote{\url{https://huggingface.co/chestnutlzj/ChatLaw-Text2Vec}.}~\cite{cui2023chatlaw} is a legal text matching model based on ChatLaw which is pre-trained on a corpus of 936,727 legal documents. 
\textbf{G-DSR}~\cite{louis2023finding} uses legal-CamemBERT\footnote{\url{https://huggingface.co/maastrichtlawtech/legal-camembert-base}.}, a legal variant of CamemBERT
%\footnote{\url{https://huggingface.co/etalab-ia/dpr-question_encoder-fr_qa-camembert}.} 
trained on BSARD dataset. It takes into account both the dense representation of text and the graph representation of legal structures. G-DSR is the state-of-the-art SAR method in the French legal domain.
\textbf{SAILER}
% \footnote{\url{https://huggingface.co/CSHaitao/SAILER_zh}.}
~\cite{sailer} is a structure-aware LCR model. It adopts an asymmetric encoder-decoder architecture to integrate structures of legal case document information into dense vectors. SAILER achieves state-of-the-art retrieval performance in Chinese LCR domain. As for the training of all baselines, we follow~\citep{DSI,NCI,DSI-QG} and use Approximate nearest neighbor Negative Contrastive Estimation (ANCE)~\cite{xiong2020approximate} method.

\textbf{(3) Generative retrieval methods:}\ \textbf{DSI}~\cite{DSI} is a new paradigm for document retrieval tasks. It utilizes a Transformer-based encoder-decoder model to map queries directly to relevant IDs. DSI achieves the end-to-end retrieval.
\textbf{NCI}~\cite{NCI} improves DSI in terms of using constrained beam search and prefix-aware weight-adaptive decoder. Both DSI and NCI use hierarchical $k$-means clustering of document vectors to create $k$-means IDs.
\textbf{DSI-QG}~\cite{DSI-QG} design a query generation process on the top of DSI, which can mitigate data distribution mismatches present between the indexing and the retrieval phases.
\textbf{Ultron}~\cite{zhou2022ultron} improves DSI through adopting product quantization to create semantic IDs and using URLs to create term-based IDs.

\subsubsection{Implementation Details}
\label{sec:app:implementation}
We implement baseline methods following the suggestions in the original papers. 
 
(1) For classical term-based baselines, we use the pyserini 
% \footnote{\url{https://github.com/castorini/pyserini}.} 
and genism
% \footnote{\url{https://github.com/RaRe-Technologies/gensim}.}
 toolkits with the default parameters. 

(2) For dense retrieval baselines, we use the open-sourced checkpoint to initialize model parameters of the pre-trained models and use faiss
% \footnote{\url{https://github.com/facebookresearch/faiss}.}
toolkit to implement ANN algorithms. The batch size is set to 16. The max length of the input text is set to 1024 for Lawformer and 512 for 
the other models. We tune these models with in-batch contrastive loss.

(3) For generative retrieval baselines, we directly use their official open-source implementations, employing the pre-trained T5 ``Randeng"\footnote{\url{https://huggingface.co/IDEA-CCNL/Randeng-T5-77M-MultiTask-Chinese}.} as backbone for Chinese legal domain, and ``t5-base''\footnote{\url{https://huggingface.co/t5-base}. Please note that this is applicable to French datasets.} for the French legal domain. We use beam search to retrieve relevant cases, where the beam size is set to 30.

We keep the backbone model and beam size same as baselines and set the max input length of GEAR to 512, the rest hyperparameters are tuned as follows:
the batch size is set to 2; the learning rate is tuned from [1e-5,1e-4] with step size 2e-5;
in the rationale extraction module, $k_1$, $k_2$, $k_3$ are respectively set to 2, 5, 15 for ELAM and 10, 20, 15 for LeCaRDv2; $\lambda_1$, $\lambda_2$, $\lambda_3$ are  respectively set to 10.0, 1.0, 0.1;
in the law structure constraint tree module, the height of the tree (the length of the hierarchical ID) $L$ is set to 4;
in the training, $\mu$ is set to 1; $\gamma_h$ is tuned from $\{0.01,0.1\}$; $\gamma_h$ is set to 1; $\lambda_w$ is tuned from $\{1,10,100\}$; $\lambda_l$ is tuned from [1e-4,1e-2] with step size 5e-4.
Hyperparameters of GEAR are tuned using grid search with Adam\citep{kingma2014adam}. ALL experiments are conducted on a single NVIDIA RTX A6000. The source code and datasets have been shared at:~\url{https://github.com/E-qin/GEAR}.
% ~\url{https://anonymous.4open.science/r/GEAR}.
\begin{table}[t]
\footnotesize
\caption{Statistics of the datasets. Since queries in BSARD do not involve judgments, we omit the judgment prediction comparison on this dataset.}
 \label{tab:dataset statistics}
    \centering
    \begin{tabular}{l|c|c|c}
\toprule
&\multicolumn{3}{c}{\textbf{Dataset}} \\
 \textbf{Statistics} & ELAM  & LeCaRDv2 & BSARD\\ 
 \midrule
 % Language & Chinese & Chinese \\ 
 
 Avg. length per candidate document & 1163.68 & 1568.38 & 880.29 \\ 
 Avg. length per query document  & 1304.98 & 558.18 & 92.48\\ 
 Avg. \# charges per candidate case  & 1.00 & 1.50 & - \\
 % Avg. \# charges per query case (Top 1)  & 1.00 & 1.53 & - \\

%  S.D. length per candidate case & 795.98 & 851.31 \\
% S.D. length per query case & 1004.58 & 527.41 \\
 \# Available candidates per query & 1332 & 1390 & 1612\\
 \# Query documents involved & 147 & 653 & 1108 \\
 \# Charges involved of judgment & 97 & 100 & - \\ 

 \bottomrule
\end{tabular}
\end{table}
% is the retrieval version of ELAM~\cite{yu2022explainable}, a publicly available dataset of legal case matching task. To simulate a more realistic retrieval scenario, we use the fact part of a real case as a query case, thus transforming the ELAM into a retrieval dataset format. Specifically, we take a number of cases from the cases involved in ELAM as query cases without judgment part, and then construct related cases as labels for each query case using ELAM's original 3-level matching labels from 2 to 0. The related cases for each query are sorted according to the relevance level, following the common format of retrieval datasets. We set the candidate pool size to 1332 for query cases.

% Queries here only contain the
% fact paragraph while the candidate documents are the entire case. \qwc{ Explain why we only use the fact in cases? or Not?}
\begin{table*}[t]
    \small
\caption{Performance comparisons of our approach and the baselines on ELAM dataset and LeCaRDv2 dataset. The best and the second-best performances are denoted in bold and underlined fonts, respectively. ``R@K'' is short for ``Recall@K''.
    $^\dagger$ denotes GEAR performs signifcantly better than baselines  based on two-tailed paired t-test with Bonferroni correction ($p < 0.05$).}
    \label{tab:Exp:matching}
\centering
    \begin{tabular}{l|ccccc|ccccc}    
        \toprule
        &\multicolumn{5}{c|}{ELAM}
        &\multicolumn{5}{c}{LeCaRDv2}
        \\
         \textbf{Models}  & \textbf{R@1} & \textbf{R@5}& \textbf{R@10} & \textbf{R@20}& \textbf{MRR} & \textbf{R@1}& \textbf{R@5} & \textbf{R@10}&\textbf{R@20}& \textbf{MRR}\\

        \midrule
        % \multicolumn{11}{c}{Sparse Retrieval Methods} \\
        % \midrule
        
        % tf-idf & 0.0090 & 0.0614 & 0.1368 & 0.2185 & 0.0662 & 0.0554 & 0.2008 & 0.3768 & 0.5449 & 0.1794 \\ 
        QL~\cite{zhai2008statistical} & 0.0272 & 0.1088 & 0.1361 & 0.2857 & 0.0723 & 0.0252 & 0.0892 & 0.1351 & 0.2177 & 0.1327 \\ 
        BM25~\cite{robertson2009probabilistic} & 0.0340 & 0.0680 & 0.1497 & 0.2245 & 0.0635 & 0.0435 & 0.1452 & 0.2549 & 0.3900 & 0.1862\\  
        % bm25+Doc2query & 0.0136 & 0.0612 & 0.1294 & 0.2391 & 0.0899 & 0.0258 & 0.1111 & 0.1932 & 0.3031 & \textcolor{blue}{0.2108}\\ 
        \midrule
        % \multicolumn{11}{c}{Dense Retrieval Methods} \\
        % \midrule
        % ANCE~\cite{xiong2020approximate} & \underline{0.0729} & 0.1894 & 0.2357 & 0.3300 & 0.1512 & 0.0362 & 0.0929 & 0.0968 & 0.1145 & 0.1323\\
        BERT~\cite{devlin2018bert} & 0.0302 & 0.1008 & 0.1405 & 0.2861 & 0.1521 & 0.0299 & 0.0718 & 0.1557 & 0.2534 & 0.1162\\
        Legal-BERT~\cite{zhong2019openclap} & 0.0384 & 0.0938 & 0.1509 & 0.3123 & 0.1510 & 0.0218 & 0.0620 & 0.1081 & 0.2743 & 0.1138\\
        Lawformer~\cite{lawformer} & 0.0537 & 0.1682 & 0.2220 & 0.3789 & 0.1701 & 0.0518 & 0.1491 & 0.2728 & 0.3593 & 0.1638\\
        ChatLaw-Text2Vec~\cite{cui2023chatlaw} &0.0385	&0.1371	&0.2065	&0.3323	& 0.1694 & 0.0356&	0.0813&	0.1510&	0.3380&	0.1379 \\
        SAILER~\cite{sailer} & \underline{0.0729} & \underline{0.2132} & \underline{0.3282} & \underline{0.4604} &\underline{ 0.2029} & \underline{0.0608} & \underline{0.1644} & \underline{0.2910} & \underline{0.4271} & \underline{0.2018}\\ 
        \midrule
        % \multicolumn{11}{c}{Generative Retrieval Methods} \\
        % \midrule
        DSI~\cite{DSI} & 0.0204 &	0.1134 & 0.2274 & 0.3159 & 0.1249 & 0.0232 & 0.0577 & 0.0768 & 0.1285 & 0.1159 \\
        DSI-QG~\cite{DSI-QG} & 0.0278 & 0.1606 & 0.2736 & 0.3803 & 0.1531 & 0.0283 & 0.0725 &	0.1230 & 0.1881 & 0.1224\\
        NCI~\cite{NCI} & 0.0325 & 0.0936 & 0.1463 & 0.2070 & 0.1285 & 0.0416 & 0.1024 & 0.1696 & 0.2504 & 0.1914 \\
        Ultron~\cite{zhou2022ultron} & 0.0607 & 0.1583 & 0.2260 & 0.3506 & 0.1678 & 0.0333 & 0.1207 & 0.2142 & 0.3492 & 0.1511 \\
        % \ywj{we need more}\\
        \midrule
        GEAR& \textbf{0.0793}$^\dagger$&	\textbf{0.2368}$^\dagger$&	\textbf{0.3356}$^\dagger$&	\textbf{0.4976}$^\dagger$&		\textbf{0.2365}$^\dagger$ & \textbf{0.0630}$^\dagger$ &	\textbf{0.1706}$^\dagger$ &	\textbf{0.3142}$^\dagger$ &	\textbf{0.4625}$^\dagger$ &	\textbf{0.2162}$^\dagger$\\

        \quad w/o $\mathcal{L}_c$&
        0.0611&	0.1657&	0.2793&	0.4167&		0.1763 & 0.0452 &	0.1485 &	0.2549 &	0.4478 &	0.1979\\

        \quad w/o $f_R$&
       0.0586& 0.1830& 0.2913& 0.4437& 0.1802 & 0.0308	& 0.1143 &	0.2108 &	0.3714 &	0.1621\\

        \quad w/o $f_T$&
       0.0464& 0.1429& 0.2328& 0.3158& 0.1626 & 0.0166 &	0.0550 &	0.0971 &	0.1388 &	0.1596  \\

        \bottomrule
    \end{tabular}
\end{table*}

% \subsubsection{Baselines} 
% \label{subsubsec:Baselines and Evaluation Metrics}
% We compare GEAR with three types of representative baselines in this work including term-based methods, dense retrieval methods, and generative retrieval methods. We leave the introduction on baselines and the detailed implementation in Appendix~\ref{sec:app:baseline} and Appendix~\ref{sec:app:implementation}, respectively.

% \szh{give a template. Need to fill the introduction of baselines.}
\subsubsection{Evaluation Metrics} 
For a fair comparison, we follow previous works~\cite{DSI, DSI-QG, NCI, zhou2022ultron} and leverage the commonly adopted metrics, including Recall (R) and MRR.
The averaged results on all test cases are reported.

To demonstrate GEAR's ability on judgment prediction, we introduce a new metric called $coverage@k$, assessing the percentage of charges involved in the query that are covered by top-$k$ retrieved documents.
This metric evaluates the charge-level consistency, i.e., the extent to which the retrieval process contributes to the efficacy of legal judgment. Formally, $coverage@k$ is defined as:
\begin{equation}
     coverage@k = \frac{1}{N_q}\sum_{i=1}^{N_q}\frac{| \mathcal{E}_{i}^{\text{Top-}k} \cap \mathcal{E}_{i} |}{|  \mathcal{E}_{i} |},
\end{equation}
where $|~|$ is the size of a set; for $i$-th testing query, $\mathcal{E}_{i}$ denotes its label set of applicable judgment charges, $\mathcal{E}_{i}^{\text{Top-}k}$ denotes the the set of charges of the top-\(k\) retrieved documents;
$N_q$ is the number of testing queries.
% Considering that query are usually undecided and LeCaRDv2 does not have the ground-truth judged charges for query cases, we adopt $coverage_\text{freq}$ and $coverage_\text{top1}$ to create a pseudo label for each query case.
% In terms of $coverage_\text{freq}$, for each query case, we create a charge dictionary from all the charges involved in all relevant candidate cases and select the charge that appears most frequently as the charge label for the query.
% In terms of $coverage_\text{top1}$, for each query case, we take the charge involved in the most relevant candidate case as the charge label for the query.
In our experiments, we compute $coverage@k$ metric at 1, 3, 5, and 10.

\subsection{Retrieval Performance}
\autoref{tab:Exp:matching} presents the retrieval performance of GEAR and baselines on the ELAM and LeCaRDv2. All the methods are trained 10 times and the averaged results are reported.
From the results, we have the following observations for \textbf{RQ1}:

% $\bullet$ \textbf{GEAR exhibits superior performance across both datasets compared to other generative retrieval models.}
% $\bullet$ 
(1) \textbf{GEAR demonstrates a significant performance advantage over all baseline methods on both datasets.}
The relative improvements of MRR on the ELAM and LeCaRDv2 are at least 16.55\% and 7.13\%, respectively. These results indicate GEAR’s effectiveness. We attribute
the improvement to the elaborate design for explicitly integrating judgment into case retrieval including rationale extraction, law-aware ID assignment, and the revision loss. See Section~\ref{sec:exp:ablation} for the detailed analysis of each module of GEAR. 
% Among these generative retrieval baselines, Ultron perform second best, it is because Ultron employs keyword-based ID, which 
% These results indicate the effectiveness of 
% Among these models, Ultron, employing semantic substrings as IDs, outperforms DSI and DSI-QG, which rely on numeric IDs. \qwc{It is $k$-means....}
% This observation suggests that the utilization of semantics in semantic IDs improves retrieval performance.
% Then, we find that GEAR with specialized law knowledge outperforms Ultron by a large margin.
% % For instance, on the ELAM dataset, GEAR outperforms Ultron by 41\% in terms of MRR.
% % This phenomenon validates that injecting law knowledge into the model can enhance the retrieval performance for legal case retrieval.
% \szh{This paragraph explains why GEAR outperforms than other generative methods}
% $\bullet$
(2) \textbf{Compared to sparse retrieval and dense retrieval methods, current generative retrieval methods struggle to achieve satisfying performance in the legal domain.}
% Generative retrieval baselines merely focus on mapping the raw query document to its $k$-means IDs
% \footnote{$k$-means IDs are obtained through hierarchical $k$-means clustering of the vectors encoded from documents.} 
% of relevant documents. 
Without injecting the law knowledge, generative retrieval baselines inevitably yield sub-optimal results. On the other hand, SAILER introduces an implicit training objective that uses the fact description of the legal document to predict its judgments. In this way, SAILER produces judgment-aware document representations and achieves the second-best performance. However, SAILER fails to provide explicit evidence of judgment consistency for relevance modeling, leading to an inferior retrieval performance compared to GEAR.

\begin{figure}
    \centering
    \includegraphics[width=0.7\linewidth]{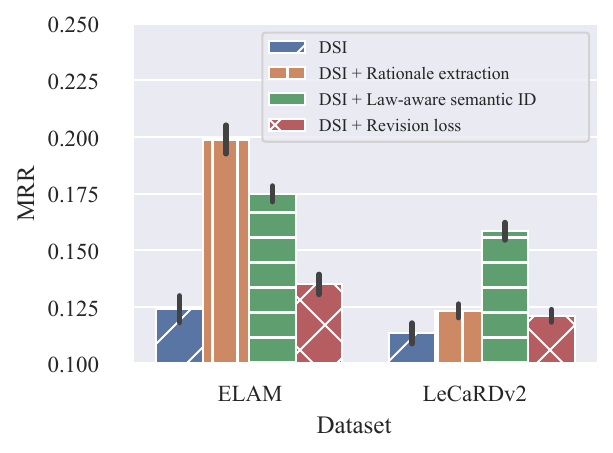}
    \caption{Comparison of retrieval performance of DSI and DSI equipped with the proposed three modules. The means values of 5 repeated experiments are reported, with error bars representing the 95\% confidence interval of the means. 
    % The performance improvements indicate the effectiveness of our components.
    }
    \label{fig:ablation}
\end{figure}

\begin{figure}
\centering
\subfigure[$coverage$ on ELAM]{
\centering
\label{fig:coverage:c}
\includegraphics[width=0.48\linewidth]{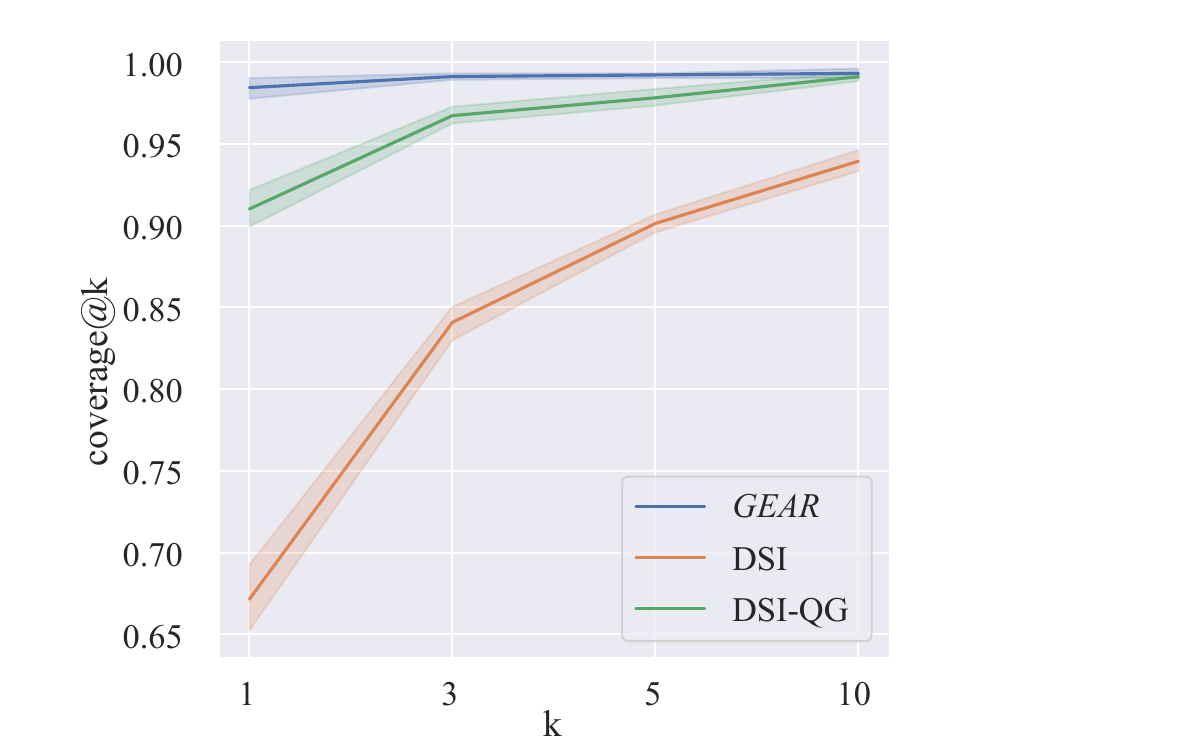}}
\hfill
\subfigure[$coverage$ on LeCaRDv2]{
\label{fig:coverage:b}
\centering
\includegraphics[width=0.48\linewidth]{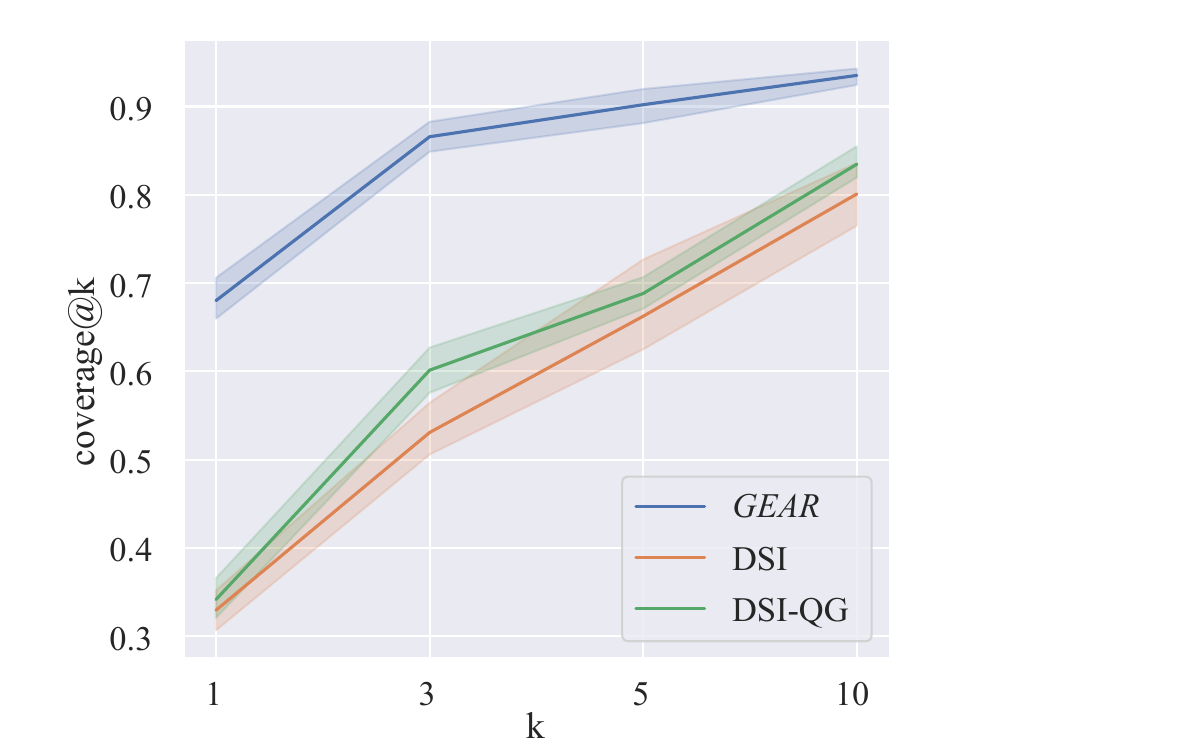}}
\caption{Comparison of the judgment prediction performances in terms of \textit{coverage}. The proposed GEAR consistently outperforms DSI and DSI-QG in both single (ELAM) and multiple (LeCaRDv2) charges scenarios.} 
\label{fig:Coverage}
\end{figure}

\subsection{Ablation Studies}
\label{sec:exp:ablation}
To answer \textbf{RQ2}, we conduct an ablation study to investigate the impact of each component of GEAR.

Firstly, we test the performance of GEAR's variants by removing a certain component. Following DSI~\citep{DSI}, we replace rationale extraction and law-aware hierarchical IDs with Direct indexing (using the first 512 tokens of document as queries for language models) and $k$-means IDs.
As shown in the bottom part of \autoref{tab:Exp:matching}, 
% The first 512 tokens of text and the $k$-means IDs obtained following DSI's settings are used for GEAR without rationale extraction and without law-aware IDs, respectively.
% % As illustrated in the last part of \autoref{tab:Exp:matching}, we conducted experiments by removing a certain component from GEAR, while all other conditions were held constant. From the results, 
we observe that: 
(1) Removing modules $\mathcal{L}_c$, $f_R$, and $f_T$ individually results in a performance degradation of 25.51\%, 23.83\%, and 31.16\% on ELAM, 8.50\%, 25.00\%, and 26.13\% on LeCaRDv2. These results verify the effectiveness of all three modules.
% The performance of GEAR with all components intact shows consistent improvement across both datasets compared to GEAR that removes one component. GEAR exceeded its residuals by 16\% to 194\% in various metrics. This suggests that all three components play some positive role in GEAR 
(2) It is worth noting that removing our law-aware hierarchical IDs results in the most significant performance decrease. It is because we inject law knowledge into each digit of the ID, aligning them with the hierarchical structure of laws. By simulating the retrieval to mirror the legal decision-making process, we enable GEAR to effectively learn the association between query cases and candidate cases. These results demonstrate the effectiveness of our IDs and highlight the importance of introducing structural and semantic knowledge in law to legal document retrieval.

% the removal of our hierarchical IDs results in the most significant performance drop. This is because we have imbued each digit of the ID with legal significance, aligning them with the legal hierarchical structure. We simulate the retrieval process to mirror the judicial decision-making process, making it easier for the model to learn the associative relationships between query cases and candidate cases."
% % It is noteworthy that the removal of law-aware IDs exerts the most substantial impact on GEAR's performance, leading to the most significant decline, which, in certain metrics, causes it to fall below the performance of other baseline models. This underscores the paramount importance of IDs in the system. Besides, revision loss is closely linked to the decoding process, and both IDs and the decoding process are governed by the law structure constraint tree. Thus, employing revision loss without the corresponding law-aware IDs result in a detrimental effect on optimization.
% Secondly,  
To further validate the efficacy of the components of GEAR, we combine the three components of GEAR with DSI individually and test its retrieval performance.
We conduct 5 repeated experiments and present the mean values of MRR, with error bars representing the 95\% confidence interval.
As illustrated in \autoref{fig:ablation},
integrating the components of GEAR into DSI consistently results in performance improvements.
% we conducted the experiment with five replications and presented the mean values of MRR, with error bars representing the 95\% confidence interval of the means. \qwc{←need check..}
% The results reveal that the incorporation of any of these components into DSI consistently results in performance improvements over the original DSI model.
Specifically, DSI equipped with rationale extraction achieves an improvement of over 60\% on ELAM, and DSI equipped with law-aware IDs demonstrated a 42\% improvement on LeCaRDv2 dataset. However, DSI equipped with revision loss does not exhibit a substantial improvement. Because DSI utilizes $k$-means IDs and the length of these IDs may vary due to differences in cluster sizes. In such cases, our revision loss cannot accurately measure the differences between predictions and labels, resulting in a slight performance improvement.
% This once again reaffirms that revision loss does not effectively operate in the absence of the corresponding ID. 

% In summary, the ablation experiments demonstrate the effectiveness of all three components of GEAR.

\subsection{Quality of Judgment Prediction}
To answer \textbf{RQ3}, we conduct experiments to evaluate the judgment prediction accuracy of retrieved cases using the proposed metric $coverage@k$. Please note that we consider the applicable charges as the judgment results. 
% Considering that in LeCaRDv2, the ground-truth judgment label of query cases is not provided, we use $coverage_\text{freq}$ and $coverage_\text{top1}$ to create pseudo labels for them.

% Considering that in LeCaRDv2, the ground-truth judgment label of query cases is not provided, we use $coverage_\text{freq}$ and $coverage_\text{top1}$ to create pseudo labels for them.
We run experiments ten times and present the results in~\autoref{fig:Coverage}, reporting the average score with the shaded area indicating the 95\% confidence interval.
From the plot, we observe that:
(1) in the single-charge scenario (ELAM) as shown in \autoref{fig:Coverage}(a)
, GEAR achieves a remarkably high $coverage$ score (over 0.95) with just 1 case retrieved. In the multi-charge scenario (LeCaRDv2) as shown in \autoref{fig:Coverage}(b), GEAR retrieves about 3 cases to encompass approximately 85\% of the charges. These results demonstrate that GEAR has considerable ability in charge prediction. 
This capability stems from our integration of judgment prediction into case retrieval, unifying the predictions for both tasks in a traversal on the law structure constraint tree. 
% In other words, we not only require the model to retrieve relevant cases but also ensure that the retrieved cases share the same applicable judgments as the query case. This integration brings the judgment prediction capability to GEAR.
(2) GEAR demonstrates a significantly \textit{coverage} improvement compared to DSI and DSI-QG on both datasets, especially when a limited number of cases are retrieved, such as 1 or 3.
This mainly attributes to the specialized design of GEAR for judgment prediction tasks including the law-aware hierarchical IDs and the revision loss, which explicitly enhances the accuracy of judgment predictions.
These results verify GEAR is capable of performing competitive legal judgment predictions.

\subsection{Effectiveness and Efficiency of Rationale Extraction}
To answer \textbf{RQ4}, we conduct experiments and evaluate the rationales extracted by GEAR
in comparison to those generated by prevalent query generation methods featured in existing generative retrieval studies. We consider the following baselines: Direct indexing~\citep{DSI}, which means using the first 512 tokens of the raw document as the query; Doc2query~\citep{DSI-QG,NCI}, which uses a language model to generate pseudo text as the model input; and ChatLaw~\citep{cui2023chatlaw}, which demonstrates to provide legal summaries that are on par with human-level quality. 

% we conducted analytical experiments on the texts generated by rationales and other methods. 
% we conducted analysis experiments on the texts from rationales extraction and that generated by other methods such as taking the first 512 tokens\footnote{The experimental setup in this paper limits the maximum length of case text to 512 tokens in Chinese.}~\citep{DSI} and pseudo text of the query generation model like Doc2query~\citep{DSI-QG,NCI, zhou2022ultron, tang2023semantic}. Since recent studies ~\citep{zhang2023benchmarking, bang2023multitask, liu2023gpteval,wang2023zero} have indicated that large language models like ChatGPT can produce text summaries of quality on par with those created by humans, we include ChatGPT summary in the scope of this comparisons to observe whether the text generated by rationales extraction is competitive with ChatGPT text and analyze the respective emphases and differences. 
First, we follow the practice~\citep{zhao2021lirex, yu2022explainable} and conduct the human evaluation to assess the quality of extracted rationales. We randomly selected 50 samples from both ELAM and LeCaRDv2 and asked two annotators (Ph.D. in Law) to determine whether the extracted rationales (generated queries) are sufficient to ascertain the applicable charges for the original cases, indicating that the extracted sentences comprehensively encompass the primary information from the original cases. Each annotator is provided with a pair of the rationales (generated queries) and the corresponding original case for each sample and is asked to mark the sample as 1 if they agreed with it and 0 otherwise. We calculate the accuracy of the sample rationales based on annotators' evaluation. As shown in \autoref{tab:human eval}, the results illustrate rationales extracted by GEAR consistently outperform both Direct indexing and Doc2query on two datasets, achieving performance comparable to that of ChatLaw. These results confirm the rationales extracted by GEAR are informative for both tasks.

% consistently outperform Direct indexing and Doc2query, and achieve on par with ChatGPT.
% achieved a competitive level of accuracy between the generated text and the original case text. On one hand, due to the rich infusion of legal knowledge, rationales extraction far outperforms the methods of using first 512 tokens or pseudo text through Doc2query.

Then, we conduct an experiment to validate the time overhead of rationale extraction.
% comparison between the time overhead of GEAR for rationale extraction and that of the baseline methods.
Since Direct indexing takes the first 512 tokens as the query, we omit its time overhead. 
% Given that the runtime of ChatGPT API may be affected by network latency, we thus conduct 50 tests and report the average time overhead and its 95\% confidence interval as shown in \autoref{tab:human eval}. 
% and we compared the times across these 50 tests. \ywj{chatGPT time} . Therefore, the network latency in the setting did not have a significant impact to results.
From the results shown in~\autoref{tab:time}, we can see that GEAR achieves impressive accuracy on par with ChatLaw but with far less time consumption. Compared to Doc2query, GEAR also exhibits a significant advantage in terms of time overhead. These results verify the GEAR's efficiency in terms of rationale extraction. Based on the results, the generative approaches for rationales are not advisable for legal document retrieval due to the efficiency issue.

\begin{table}
\footnotesize
\caption{Time overhead and human evaluation of rationale extraction over 50 random samples from ELAM and LeCaRDv2 by two annotators with the inter-rater agreement of 0.96. ``Time''(ms) denotes the extraction time per sample.}
\label{tab:human eval}
    \centering
    \begin{tabular}{l|l|c|l|c}
\toprule
&\multicolumn{2}{c|}{ELAM} & \multicolumn{2}{c}{LeCaRDv2} \\
 \textbf{Methods}& \textbf{Acc.} & \textbf{Time} & \textbf{Acc.} & \textbf{Time}\\ 
 \midrule
 Direct indexing~\citep{DSI} & 0.66 & - & 0.38 & - \\
 Doc2query~\citep{DSI-QG} & 0.48 & 4094.78(±1363.07) & 0.26 & 3947.01(±1229.81)\\ 
 % ChatGPT & 0.90 & 16332.84(±6201.80) & 0.82 & 15750.02(±4853.38)\\ 
 ChatLaw~\citep{cui2023chatlaw} & 0.94 & 20975.92(+3858.29) & 0.88 & 21827.05(±5820.58)\\
 \midrule
 $f_R$ in GEAR & 0.94 & \textbf{0.10(±0.01)} & 0.86 & \textbf{0.67(±0.03)}\\
 \bottomrule
\end{tabular}
% \vspace{-2ex}
\end{table}

\begin{table}
\small
\caption{Comparison in average inference wall time (95\% confidence interval) per query on a single NVIDIA RTX A6000.}
\label{tab:time}
    \centering
    \begin{tabular}{l|c|c}
\toprule
&\multicolumn{2}{c}{\textbf{Dataset}} \\
 \textbf{Models} & ELAM  & LeCaRDv2\\ 
 \midrule
 Ultron & 110.544(±24.703) & 31.854(±7.766) \\
 DSI-QG & 73.147(±17.062) & 26.369(±5.973) \\ 
DSI  & 72.646(±14.820) & 27.023(±6.084) \\ 
 \midrule
 GEAR &\textbf{59.184(±8.737)} & \textbf{20.391(±4.064)} \\
 \bottomrule
\end{tabular}
% \vspace{-2ex}
\end{table}

\begin{table}[tbp]
    \footnotesize
\caption{Performance comparisons of our approach and the baselines on BSARD.  ``R@K'' is short for ``Recall@K''.
    $^\dagger$ indicates that improvements are significant based on two-tailed paired t-test with Bonferroni correction ($p < 0.05$).}
    \label{tab:Exp:generalization}
\centering
    \begin{tabular}{l|ccccc}    
        \toprule
         \textbf{Models}  & \textbf{R@5}& \textbf{R@10} & \textbf{R@20}& \textbf{R@30} & \textbf{MRR} \\
        \midrule
        QL~\cite{zhai2008statistical} &   0.1752 & 0.2128 & 0.2698 & 0.2923 & 0.1770 \\ 
        BM25~\cite{robertson2009probabilistic} &  0.1815 & 0.2381 & 0.2871 & 0.3056 & 0.1786 \\ 
        % ANCE~\cite{xiong2020approximate}&  0.1232 & 0.1918 & 0.2319 & 0.2562 & 0.1258 \\ 
        G-DSR~\cite{louis2023finding} &  0.1636 & 0.3426 & 0.5081 & 0.6720 & 0.3012 \\
        DSI~\cite{DSI} & 	0.4578 &	0.5536 &	0.6315 &	0.6562 &	0.3955 \\
        \midrule
        GEAR& 	\textbf{0.5297}$^\dagger$&	\textbf{0.6164}$^\dagger$&	\textbf{0.7031}$^\dagger$&		\textbf{0.7170}$^\dagger$ & \textbf{0.4081}$^\dagger$ \\
        \bottomrule
    \end{tabular}
\end{table}

\subsection{Inference Time and Convergence}
\label{sec:app:time}
To answer \textbf{RQ5}, we record the inference wall time (in milliseconds) per query of baseline methods and GEAR.
This comparison is conducted on a single NVIDIA RTX A6000 across two datasets. From the results  shown in~\autoref{tab:time}, we observe that 
GEAR demonstrates superior efficiency, with an inference time of 59.184 ms (±8.737) for the ELAM dataset, and 20.391 ms (±4.064) for LeCaRDv2.
Specifically, when compared to Ultron, GEAR achieves a remarkable 46.48\% reduction in inference time for the ELAM dataset and a 36.03\% reduction for LeCaRDv2.
This is mainly because Ultron uses product quantization to create the IDs for documents. For all document embeddings, Ultron first divides embedding space into several groups and then performs $k$-means clustering on each group. It usually leads to excessively long IDs.
In the case of DSI and DSI-QG, GEAR exhibits a substantial 19.01\% and 18.59\% improvement for ELAM and a 22.64\% and 18.59\% improvement for LeCaRDv2. In DSI and DSI-QG, the tree constructed by $k$-means may be unbalanced, meaning that the lengths of case IDs are unequal. Some case IDs may be longer, which impairs the inference performance.
% These findings indicate the remarkable efficiency of GEAR.

\begin{figure}[t]
    \centering
    \includegraphics[width=0.65\linewidth]{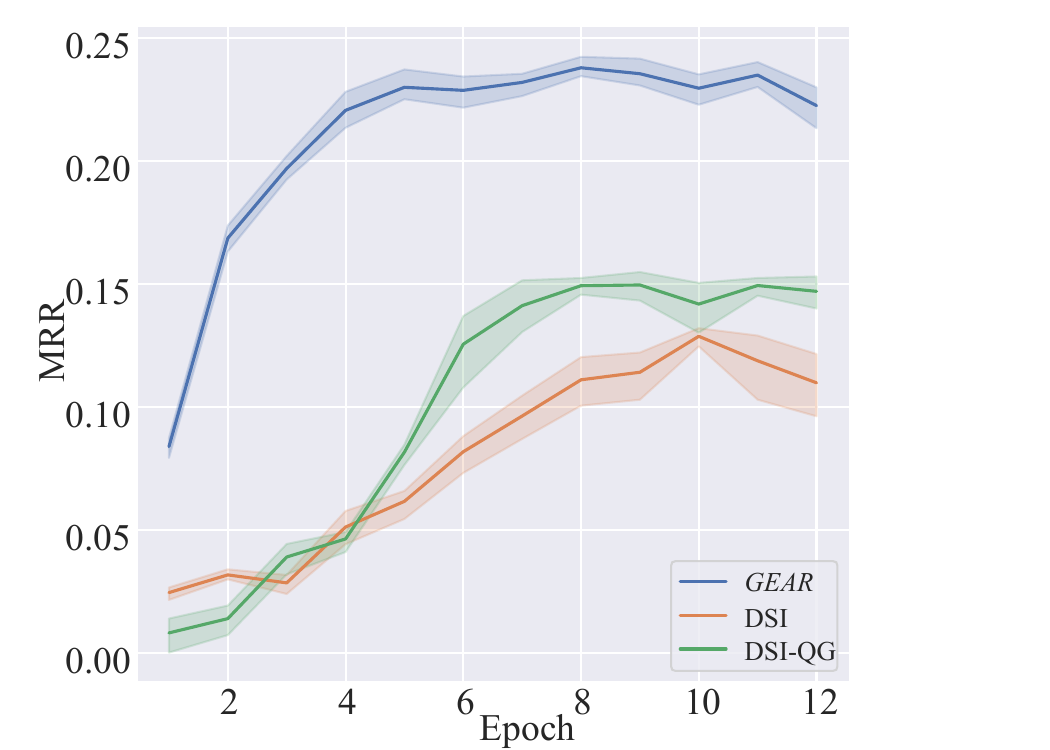}
     \caption{Testing curves of DSI, DSI-QG, and our GEAR. 
     % GEAR converges in fewer epochs (about 5 epoch), indicating its efficiency.
    }
    \label{fig:enter-label}
% \vspace{-2ex}
\end{figure}

For further confirming GEAR's efficiency, we plot the testing curves for DSI, DSI-QG, and our GEAR with the $x$-axis denoting the number of epochs, the $y$-axis denoting the MRR score, and the shaded area indicating the 95\% confidence interval. As illustrated in~\autoref{fig:enter-label}, we
 observe that GEAR not only outperforms DSI and DSI-QG significantly in terms of performance but also exhibits superior convergence. GEAR achieves near-optimal MRR performance by 6 epochs, whereas DSI-QG and DSI converge at 8 and 10 epochs, respectively. The results verify the efficiency of GEAR.

\subsection{Robustness across Languages and Domains}
\label{sec:Exp:generalization}
To answer \textbf{RQ6}, we compared the performance of baselines and GEAR on BSARD, a French SAR dataset. 
Since 
% we do not have the expertise in Belgium legal knowledge and
BSARD has a relatively short query length, we omit the rationale extraction part and take the raw legal questions as queries for GEAR. In terms of the ID, we follow the structure of the Belgian code provided in the data and assign the hierarchical semantic to each statutory article. 
% % Due to the absence of the rationales~\citep{yu2022explainable,sun2023explainable} (key circumstances and key elements~\citep{lecard,shao2023understanding}) in BSARD documents and a lack of familiarity with the complicated Belgian legal code expertise involved in BSARD, we solely construct the law structure constraint tree based on explicitly provided structural annotations of laws without extracting rationales for the documents, keeping other experimental settings unchanged. 
% \qwcC{Should we explain that the structural information of another legal system can also be utilized?}
From the results shown in \autoref{tab:Exp:generalization}, we have two observations: 
(1) with a small number of documents to retrieve, generative retrieval methods (DSI and GEAR) exhibit significantly higher retrieval performances in SAR scenario compared to dense retrieval.
We assume the reason why dense models perform poorer is that 
there exists a significant gap between legal questions and statutory articles.
It is difficult for dense models to learn the correct association between them especially without the law knowledge injected.
(2) GEAR demonstrates the best performance, exhibiting a substantial advantage over sparse and dense retrieval methods including the current state-of-the-art model G-DSR. 
The improvement of GEAR benefits from explicitly injecting legal knowledge into generative retrieval frameworks.
% Benefited by the generative retrieval framework,
% GEAR explicitly utilizes the intrinsic structural and textual knowledge of the laws, thus performing the best.
% In conclusion, GEAR exhibits strong competitiveness in statutory article retrieval, demonstrating great generalization capabilities in legal document retrieval across different languages.

% After analysis, we found that due to the significant differences in the form and length of the qurey and candidate in SAR, there might exist a certain gap between them even after mapping to dense vectors in high-dimensional space.  Generative retrieval avoids this step, so it may be more suitable for this scenario than dense retrieval. Within the framework of generative retrieval, GEAR fully utilizes the intrinsic structural and textual knowledge of the laws, thus performing the best.

\section{Conclusion}
In this study, we introduce GEAR, a novel law-guided generative legal document retrieval method that explicitly integrates judgment prediction. GEAR exploits the law knowledge and extracts rationales from legal documents, ensuring a shared and informative representation for both tasks. Grounded in the inherent hierarchy of laws, GEAR constructs a law structure constraint tree and assigns the law-aware semantic ID to each document. These designs enable a unified traversal from the root, through intermediate charge nodes, to case-specific leaf nodes, which empowers GEAR to perform dual predictions for judgment and
relevant documents in a single inference. With the help of the proposed revision loss, GEAR jointly minimizes the discrepancy between the IDs of predicted and labeled judgments/ retrieved documents, improving the accuracy and consistency for both tasks.
% Extensive experiments on two Chinese and a French datasets demonstrate that GEAR consistently outperforms
% state-of-the-art methods in legal document retrieval while maintaining competitive judgment prediction performance.
Extensive experiments on two LCR datasets show the superiority of GEAR over state-of-the-art methods while maintaining competitive judgment prediction performance. 
Moreover, we validate its robustness across languages and domains on a French SAR dataset.
% Moreover, we validate
% the effectiveness of GEAR on a French statutory article retrieval dataset, reaffirming its robustness across languages and domains.

\begin{acks}
This work was funded by the National Key R\&D Program of China (2023YFA1008704), the National Natural Science Foundation of China (No. 62376275, No. 62377044), Beijing Key Laboratory of Big Data Management and Analysis Methods, Major Innovation \& Planning Interdisciplinary Platform for the ``Double-First Class” Initiative, PCC@RUC, funds for building world-class universities (disciplines) of Renmin University of China. 
\end{acks}

\bibliographystyle{ACM-Reference-Format}
\balance
\bibliography{sample-sigconf}

\end{document}